\documentclass[%
 reprint,
 showkeys,
 amsmath,amssymb,
 aps,
]{revtex4-2}

\usepackage{graphicx}
\usepackage{dcolumn}
\usepackage{bm}

\usepackage{xcolor} 
\usepackage{graphicx}
\usepackage{hyperref}
\usepackage{siunitx} 
\DeclareSIUnit\bar{bar}
\usepackage{subcaption} 
\usepackage{booktabs} 
\usepackage{multirow} 
\usepackage{amsmath} 
\usepackage{amsfonts} 

\begin{document}

\preprint{APS/123-QED}

\title{Spatial characterization of debris ejection from the interaction \\ of a tightly focused \si{\peta\watt}-laser pulse with metal targets}

\author{I.-M.~Vladisavlevici$^{1}$, C.~Vlachos$^{2,3,4}$, J.-L.~Dubois$^{5,2}$, A.~Huerta$^1$, S.~Agarwal$^{6,7}$, H.~Ahmed$^8$, J.~I.~Api\~{n}aniz$^1$, M.~Cernaianu$^9$, M.~Gugiu$^9$, M.~Krupka$^{6, 10}$, R.~Lera$^1$, A.~Morabito$^1$, D.~Sangwan$^9$, D.~Ursescu$^9$, A.~Curcio$^{1, 12}$, N.~Fefeu$^2$, J.~A.~P\'{e}rez-Hern\'{a}ndez$^1$, T.~Vacek$^2$, P.~Vicente$^1$, N.~Woolsey$^{11}$, G.~Gatti$^1$, M.~D. Rodr\'iguez-Fr\'ias$^{1}$, J.~J.~Santos$^2$, P.~W.~Bradford$^2$, M.~Ehret$^{1,}$}

\email{corresponding author: mehret@clpu.es}

\address{$^1$Centro de L\'{a}seres Pulsados (CLPU), 37185, Villamayor, Spain}
\address{$^2$Univ. Bordeaux - CNRS - CEA, Centre Lasers Intenses et Applications (CELIA), UMR 5107, Talence, France}
\address{$^3$ Institute of Plasma Physics and Lasers, University Research and Innovation Centre, Hellenic Mediterranean University, Rethymno, Greece}
\address{$^4$CEA, DAM, DIF, Arpajon, France}
\address{$^5$CEA, DAM, CESTA, F-33116 Le Barp, France}
\address{$^6$FZU-Institute of Physics of Czech Academy of Sciences, Prague, Czech Republic}
\address{$^7$Faculty of Mathematics and Physics, Charles University, Prague, Czech Republic}
\address{$^8$Central Laser Facility, Rutherford Appleton Laboratory, Didcot, United Kingdom}
\address{$^9$Extreme Light Infrastructure (ELI-NP) and Horia Hulubei National Institute for R~\&~D in Physics and Nuclear Engineering (IFIN-HH), M\u{a}gurele, Romania}
\address{$^{10}$Institute of Plasma Physics of Czech Academy of Sciences, Prague, Czech Republic}
\address{$^{11}$York Plasma Institute, School of Physics, Engineering and Technology, University of York, York, United Kingdom}
\address{$^{12}$Istituto Nazionale di Fisica Nucleare - Laboratori Nazionali di Frascati, Frascati (Rome), Italy}

\date{\today}

\begin{abstract}
We present a novel scheme for rapid quantitative analysis of debris generated during experiments with solid targets following relativistic laser-plasma interaction at high-power laser facilities. Experimental data indicates that predictions by available modeling for non-mass-limited targets are reasonable, with debris on the order of hundreds \si{\micro\gram}-per-shot. We detect for the first time that several \si{\percent} of the debris is ejected directional following the target normal (rear-  and interaction side); and confirm previous work that found the debris ejection in direction of the interaction side to be larger than on the side of the target rear.
\end{abstract}


\maketitle


\begin{figure*}[!htb]
    \centering
    \includegraphics[width=\textwidth]{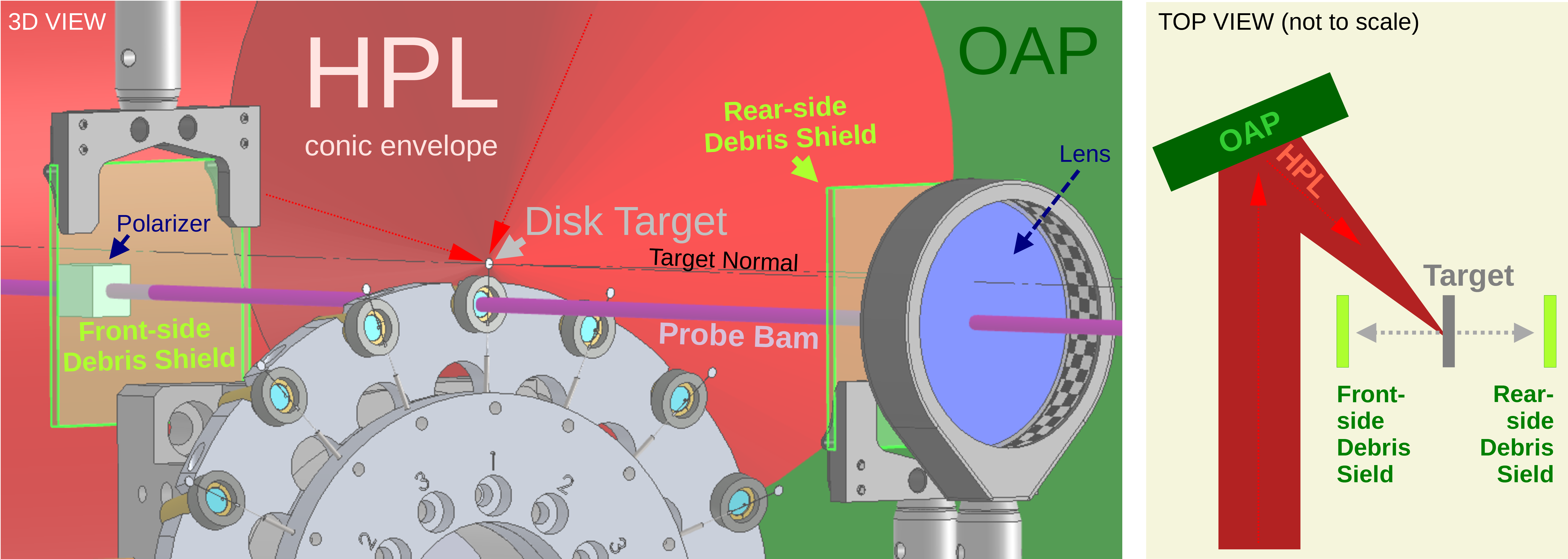}
    \caption{Two sputter plates from fused silica are used to shield probe beam optics from debris in solid-target experiments at the ELI-NP high-power laser (HPL) facility. Note the laser is focused to relativistic intensities via an off-axis parabola (OAP) onto a disk target. The front-side debris shield protects a polarizer towards the target normal on the laser-interaction side of a disk target; the rear-side debris shield catches debris in front of an imaging lens. The target normal is collinear with the normal of both debris shields.}
    \label{fig:setup-NP}
\end{figure*}

\section{Introduction}

Established high-power Ti:Sa laser systems \cite{Jeong2014, Hooker2008, Yu2012, Leemans2013, Sung2017, Li2018, Lureau2020, Danson2015, Wang2017} are able to deliver laser pulses up to several \si{\peta\watt} at a high-repetition-rate of \SIrange{0.05}{1}{\hertz}. Focusing them to relativistic intensities allows to create laser driven secondary sources in a wide range from ionizing radiation \cite{Daido2012, Borghesi2019, TajimaMalka2020, Norreys1999} to XUV- and \si{\tera\hertz}-pulses \cite{Lichters1996, Corkum2007, Beg1997, Bleko2016}. Solid density metal targets are being used to create ion sources \cite{Bin2022, Xu2023}, flashes of high energetic X-rays \cite{Steven1993, Jiang2002} and extreme ultraviolet light sources \cite{Fujioka2009}. A high-repetition-rate operation is important for many applications in medicine and fusion science \cite{Roth2001, Ledingham2004}, but poses a challenge for system integrity.

Debris management is an important aspect of ultrahigh intensity laser-solid interaction at high-repetition-rate. The amount of ejected mass ranges in the order of hundreds of \si{\micro\gram} per laser shot \cite{Ehret2023} and the deposition of the ablated material is observed to deteriorate beamline components \cite{Fujioka2009, Booth2018}. Available detailed characterizations of debris have been limited to non-relativistic laser intensities just above the ionization threshold \cite{Ando2008,Fujioka2009,Amano2010}. First characterization attempts for relativistic high-power laser interactions show a timeline of small-most debris particles ejected earlier, with a fast ejection speed, and successively larger projectiles with lower velocity \cite{Booth2018}. These studies further indicate an asymmetry of the ejection for early times, with more debris being ejected away from the side on which the laser interaction takes place, but lack a characterization of the spatially resolved debris deposition.

This paper presents a characterization of ejected debris with spatial resolution, for the first time, that will allow an evaluation of mitigation strategies to avoid damage and deterioration of beamline components, diagnostics and metrology devices.

The paper is structured as follows: (i) after a brief introduction of the novel methodology that is used to derive spatially resolved measurements from flatbed scans in Sec.~\ref{sec:materials}, (ii) we present results from an experimental campaign at a high-power laser in Sec.~\ref{sec:results} that show two distinct types of debris, and (iii) close with discussion and conclusion in Sec.~\ref{sec:discussion} and Sec.~\ref{sec:conclusion}, evaluating the amount of ejected debris and relating results to available modeling.

\section{Materials and Methods}\label{sec:materials}

Experiments for this work are conducted at the Extreme Light Infrastructure Nuclear Physics (ELI-NP) \cite{Doria2020} with a high-power \SI{1}{\peta\watt} Ti:Sa laser delivering on target $E_\mathrm{L}\approx\SI{ 22}{\joule}$ within a pulse duration of $\tau_\mathrm{L}\approx\SI{30}{\femto\second}$. The energy is extrapolated from calibrations recorded at low-energy and the pulse duration is measured on-shot with a FROG system that diagnoses a picked-up reflection from a small elliptical mirror positioned before the focusing parabola. The laser pulse is focused with an incidence angle of \SI{45}{\degree} onto \SI{50 \pm 5}{\micro\metre} thick nickel disk targets, with a focal spot diameter of $d_\mathrm{L}~\approx~\SI{4}{\micro\metre}$ full-width at half-maximum (FWHM). The focal spot at high energy is estimated to be the same as for low-energy measurements. The setup is shown in Fig.~\ref{fig:setup-NP}, with the focusing parabola (OAP) in the back and targets mounted on a wheel.

Two \SI{1}{\milli\metre} thick and $\SI{50}{\milli\metre} \times \SI{50}{\milli\metre}$ squared sputter plates from fused silica are used to catch debris that is emitted away from the respective target front and rear sides. The front sided sputter plate is placed in front of a polarizer facing the laser-interaction side, while the rear sided sputter plate is placed in front of an imaging lens. Note the auxiliary character of this arrangement of catchers as the OAP is by default protected with a thin pellicle. The plates' centres are are not perfectly collinear with the laser-interaction point, but shifted by \SI{3}{\milli\metre} down with respect to the target, and their surfaces are parallel to the target surface. The distance of the rear plate to the interaction point is \SI{125 \pm 5}{\milli\metre}, the front plate is positioned at \SI{95 \pm 10}{\milli\metre}.

\begin{figure}[!h]
    \centering
   \includegraphics[width=\columnwidth,trim=0.3cm 0.4cm 0.4cm 0.4cm,clip]{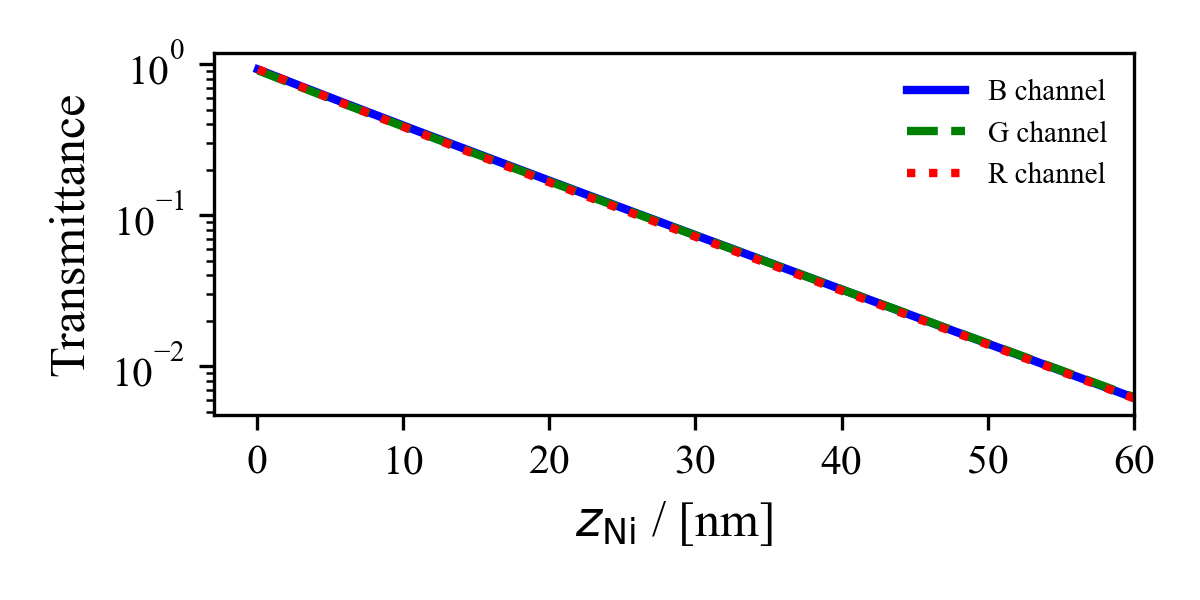}
    \caption{Predicted transmittance through nickel deposit of thickness $z_\mathrm{Ni}$ on a \SI{1}{\milli\metre} thick silica plate for three channels of a RGB scan with the EPSON V-750-PRO flatbed scanner.}
    \label{fig:transmittanceTHEO}
\end{figure}

After the experiment, the sputter plates are scanned with an EPSON V-750-PRO flatbed scanner to obtain the spatially resolved deposited debris thickness $z_\mathrm{Ni}$ as a function of the transmittance ${T} = I_\mathrm{t} / I_0$. Here $I_0$ is the intensity of the incident wave and $I_\mathrm{t}$ the intensity at the exit of the double layer system. The details of the scanning procedure are outlined in App.~\ref{app:scans} and the calculation of the transmission of a flat double-layer system is revisited in App.~\ref{app:transmission}. The theoretically predicted transmittance of evaporated nickel deposit is shown in Fig.~\ref{fig:transmittanceTHEO} as a function of the layer thickness for three channels of a color scan. One notes the good agreement between the different color channels which points to a flat spectral response.

\section{Results}\label{sec:results}

\begin{figure}[!tb]
\begin{subfigure}[c]{0.48\columnwidth}
    \centering
    \includegraphics[width=\textwidth,angle=180,trim=0.5cm 0.4cm 0.3cm 0.4cm,clip]{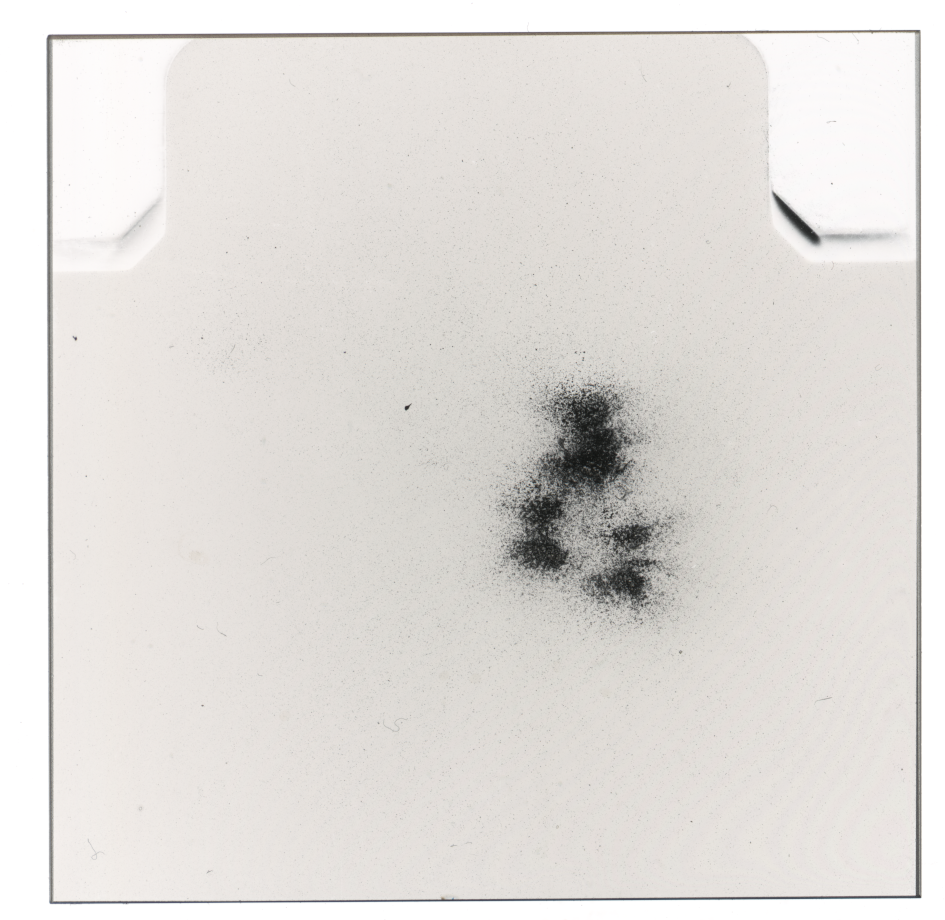}
    \vspace*{-7ex}
    \begin{center}
    (a)
    \end{center}
    \vspace*{9ex}
\end{subfigure}
\hfill
\begin{subfigure}[c]{0.48\columnwidth}
    \centering
    \includegraphics[width=\textwidth,trim=0.5cm 0.4cm 0.3cm 0.4cm,clip]{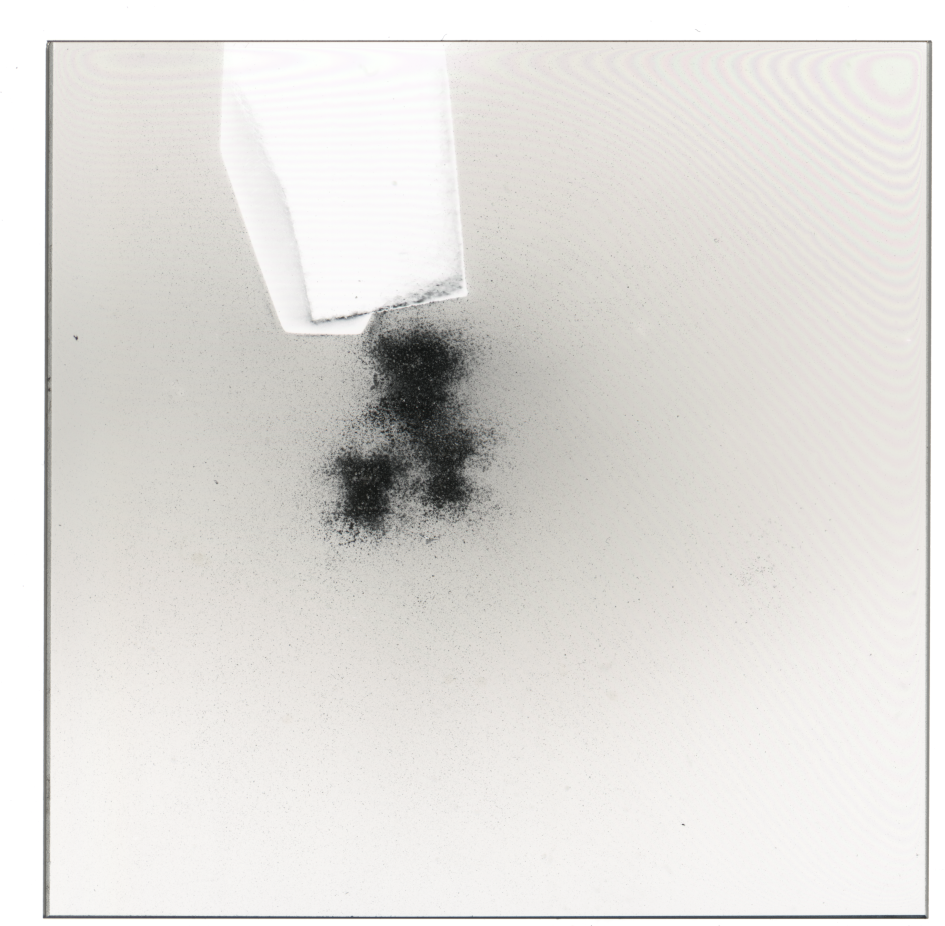}
    \vspace*{-10ex}
    \begin{center}
    (b)
    \end{center}
    \vspace*{8ex}
\end{subfigure}
    \vspace*{-10ex}
    \caption{Debris deposited on silica plates positioned in target normal direction (a) atop the target rear, and (b) atop the target front side facing the high-power laser at ELI-NP \SI{1}{\peta\watt}. The plates are \SI{50}{\milli\metre} squares.}
        \label{fig:DebrisELI-NP}
\end{figure}

Raw scans of debris collected on the sputter plates are shown in Fig.~\ref{fig:DebrisELI-NP}, with both plates having recessed areas that were protected from debris by mounting structures. Debris originates from three laser shots, two shots on targets with diameter $d_\mathrm{t} = \SI{0.5}{\milli\metre}$ and one shot on a disk of \SI{2}{\milli\metre} diameter. The sputter plates are uniformly coated by a surfacic deposition of debris, a weak but distinct areal deposition that uniformizes towards the edges of the plates. Additionally, three distinct deposition marks are observed towards the target normal. Slight target misalignment \SI{< 5}{\degree} might be responsible for the spatial separation of the marks. This hypothesis is supported by the diametrical opposition of structurally similar marks with respect to the target position.

Further, one notes two small marks contrasting one large mark and it is reasonable to assume that small marks correspond to shots on small disk targets.

\begin{figure}[!b]
    \centering
    \includegraphics[width=\columnwidth,trim=0.3cm 0.4cm 0.4cm 0.4cm,clip]{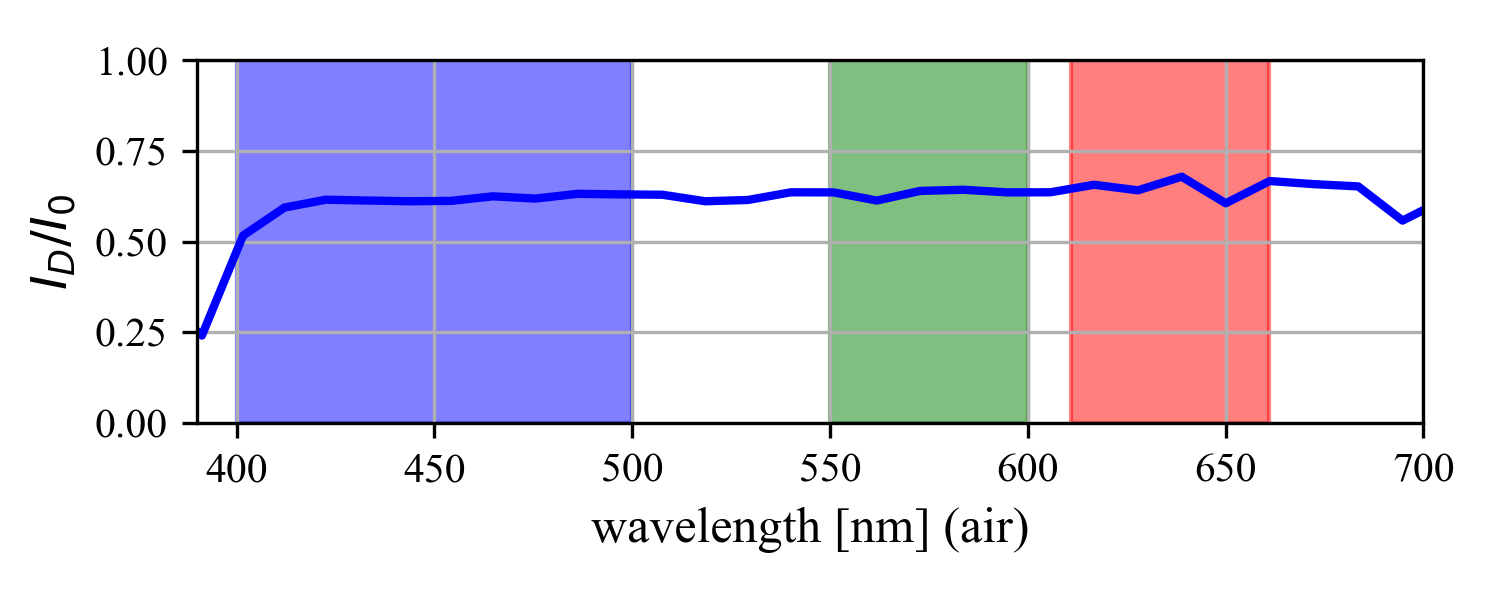}
    \caption{Spectrally resolved transmittance of nickel debris illuminated with the light source in an EPSON V-750-PRO flatbed scanner; indicated are blue, green and red bands of acquisition for the scanner head. A measurement of intensities $I_0$ through silica glass is used to normalize the measurement through debris $I_D$.}
    \label{fig:ournickel}
\end{figure}

We measured the transmittance through the centre of the rear-side sputter plate with spectral resolution using a compact Czerny-Turner spectrometer, as shown in Fig.~\ref{fig:ournickel}. The integrated surface element is \SI{1}{\square\milli\metre}. The measurement shows a flat spectral response in accordance with the theoretical prediction presented in Sec.~\ref{sec:materials} for the transition metal nickel deposited on a fused silica plate.

\begin{figure*}[!tbp]
    \centering
    \includegraphics[width=\textwidth,trim=0.4cm 0.4cm 0.4cm 0.4cm,clip]{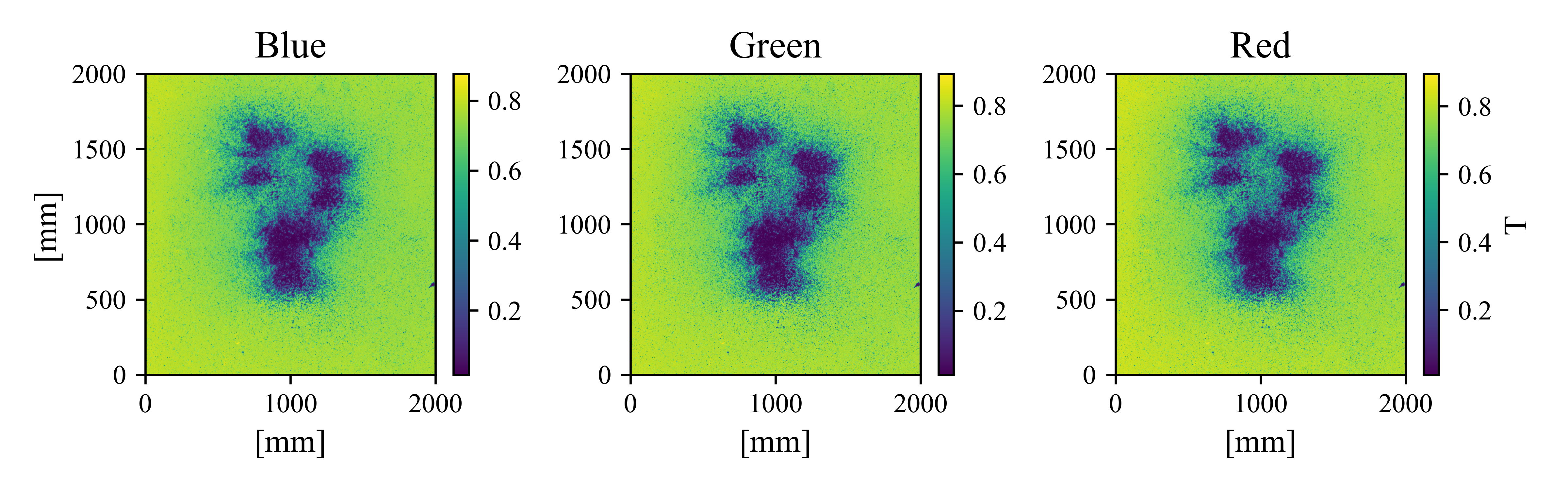}
    \caption{Transmittance through the debris on the rear-side (with respect to the laser interaction) silica plate for all three channels of the RGB scan.}
    \label{fig:transmittanceELI-NP}
\end{figure*}

The transmittance of the debris in Fig.~\ref{fig:DebrisELI-NP}~(a) is shown in Fig.~\ref{fig:transmittanceELI-NP} in a squared region-of-interest (ROI) around the area of maximum deposition. For conversion from scan intensity to transmittance we follow Eq.~\ref{eq:Trasnmittance} from App.~\ref{app:scans}, which reads
\begin{equation}
     T_{mn} = 10^{ \left( \ln{\left[ I_{mn} \right]} / B \right) - C } \quad .
\end{equation}
The raw data is analyzed separately for the distinct RGB channels, revealing no opaque zones which allows for quantitative analysis of the full surface. For all three channels the transmittance shows a similar behaviour, as can be seen in Tab.~\ref{tab:transmittance}. The surfacic deposition on the rear-side has a uniform mean maximum transmittance of $89^{+11}_{-14}\,\%$ and the deposition marks have a minimum transmittance of \SI{1.5 \pm 0.2}{\percent} in the largest spot. Towards the front side, the maximum transmittance amounts to $88^{+12}_{-19}\,\%$ and the deposition marks have a minimum transmittance of \SI{1.7 \pm 0.2}{\percent} in the largest spot.

\begin{table}[!b]
    \centering
    \caption{Minimum (min) and maximum (max) transmittance values for the front- and rear-side (with respect to the laser interaction) sputter plates across the three color channels of a RGB scan.     \label{tab:transmittance}}
\begin{ruledtabular}
    \begin{tabular}{cccc}
                               &     & front side & rear side \\ 
         \midrule
         \multirow{2}{*}{Blue} & min & \SI{2.1}{\percent}  & \SI{1.9}{\percent} \\
                               & max & \SI{86.3}{\percent} & \SI{87.6}{\percent} \\ 
         \multirow{2}{*}{Green}& min & \SI{1.5}{\percent}  & \SI{1.4}{\percent} \\
                               & max & \SI{88.2}{\percent} & \SI{89.3}{\percent} \\ 
         \multirow{2}{*}{Red}  & min & \SI{1.4}{\percent}  & \SI{1.3}{\percent} \\
                               & max & \SI{88.1}{\percent} & \SI{89.5}{\percent} \\
    \end{tabular}
\end{ruledtabular}
\end{table}

\begin{figure*}[!htb]
    \centering
    \includegraphics[width=\textwidth,trim=0.4cm 0.4cm 0.4cm 0.4cm,clip]{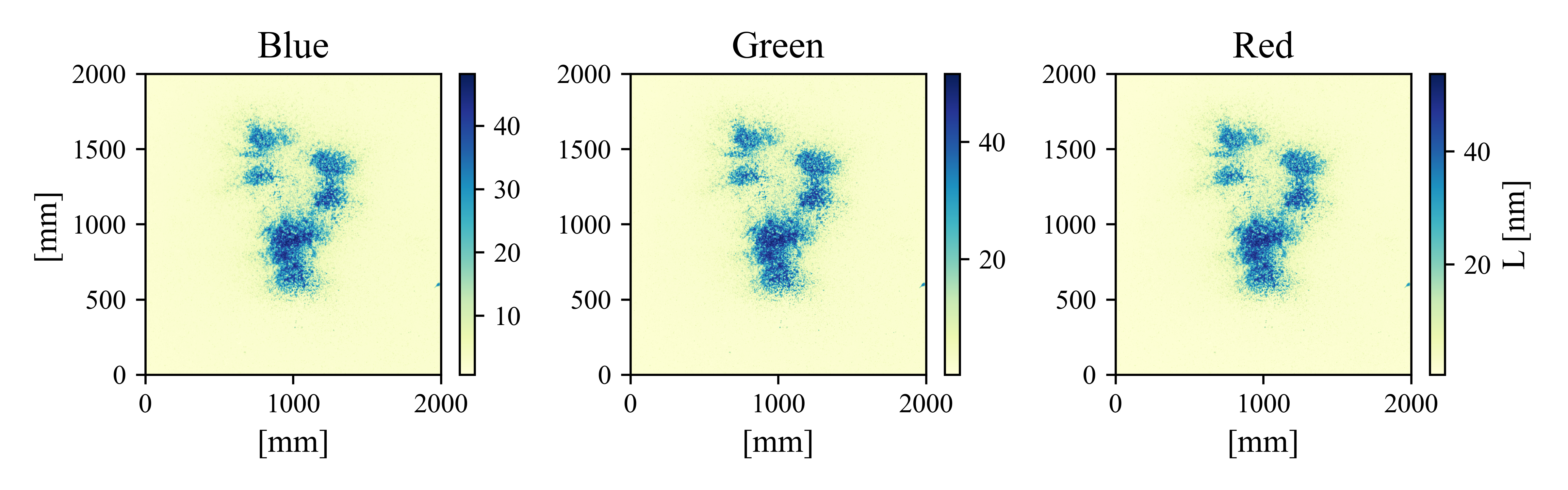}
    \caption{Thickness of the nickel debris on the rear side silica plate calculated from the transmittance separately for all three channels of the RGB scan.}
    \label{fig:thicknessdebrisELINP}
\end{figure*}

Considering that the debris on the sputter plates consists only of deposited nickel, we calculate the thickness of the deposited debris using Eq.~\ref{eq:transmittance} from App.~\ref{app:transmission}, which reads 
\begin{equation}
    T(z_\textrm{M},z_\textrm{S}) = \int_{\omega_-}^{\omega_+} \lVert \mathfrak{T}(z_\textrm{M},z_\textrm{S},\omega) \rVert^2 \cdot s(\omega) ~ \mathrm{d}\omega \quad .
\end{equation}
The results are shown in Fig.~\ref{fig:thicknessdebrisELINP}, and the characteristic values are given in Tab.~\ref{tab:thickness}. The surfacic deposition on the rear-side has a uniform mean minimum thickness of \SI{0.5 \pm 0.1}{\nano\metre} and the deposition marks have a mean maximum thickness of \SI{49 \pm 9}{\nano\metre} in the largest spot. Towards the front side, the mean minimum thickness amounts to \SI{0.6 \pm 0.1}{\nano\metre} and the deposition marks have a maximum thickness of \SI{48 \pm 9}{\nano\metre} in the largest spot.
\begin{table}[!b]
    \centering
    \caption{Characteristic minimum (min) and maximum (max) thickness values deduced from the transmittance for the front- and rear-side sputter plates across the three color channels of a RGB scan.     \label{tab:thickness}}
    \begin{ruledtabular}
    \begin{tabular}{cccc}
                               &     & front side & rear side \\ 
         \midrule
         \multirow{2}{*}{Blue} & min & \SI{0.9}{\nano\metre} & \SI{0.7}{\nano\metre} \\
                               & max & \SI{45.2}{\nano\metre} & \SI{46.4}{\nano\metre} \\ 
         \multirow{2}{*}{Green}& min & \SI{0.4}{\nano\metre} & \SI{0.3}{\nano\metre} \\
                               & max & \SI{49.3}{\nano\metre} & \SI{50.2}{\nano\metre} \\ 
         \multirow{2}{*}{Red}  & min & \SI{0.6}{\nano\metre} & \SI{0.5}{\nano\metre} \\
                               & max & \SI{49.9}{\nano\metre} & \SI{50.8}{\nano\metre} \\
    \end{tabular}
    \end{ruledtabular}
\end{table}

The mass can be calculated as $z_\mathrm{Ni} \cdot p^2 \cdot \rho $ with the pixel size $p = \SI{10.6}{\micro\metre}$ and assuming solid density $\rho = \SI{8.9}{\gram\per\cubic\centi\metre}$. The total mass of nickel deposited on both plates amounts to \SI{50 \pm 14}{\micro\gram} within the ROIs, \SI{22 \pm 4}{\micro\gram} towards the rear side and \SI{28 \pm 6}{\micro\gram} towards the front side. The average uniform surfacic deposition is \SI{49 \pm 10}{\micro\gram\per\steradian} towards the front side and \SI{70 \pm 13}{\micro\gram\per\steradian} towards the rear side. Note the non-linear relationship between measured transmittance and derived debris thickness. After subtraction of the mean surfacic deposition, the deposition marks towards the front side contain \SI{14 \pm 4}{\micro\gram} and towards the rear side \SI{10 \pm 3}{\micro\gram}.


\section{Discussion}\label{sec:discussion}

Available modelling \cite{Ehret2023} suggests that shots on small disk targets emit more debris than shots on large disk targets. The prediction for small disks is the total emission of \SI{257 \pm 50}{\micro\gram} and \SI{99 \pm 20}{\micro\gram} for large disks. This difference between large disks and small disks can be explained by a larger fraction of the laser-heated electrons held back by stronger fields in the case of smaller targets. With a larger refluxing cloud of near-relativistic electrons there are more electrons available to transfer heat to the bulk material. The model applies to cases where the evaporating mass is not limited by the available target mass (here \SI{>429}{\micro\metre} diameter disks), and where the target sizes are smaller than the maximum expansion of the laser-generated target potential during electron discharge (here \SI{<12}{\milli\metre} disks).

Experimentally, the total mass of the debris can be extrapolated from the mean surfacic deposition of \SI{60}{\micro\gram\per\steradian} assuming a spherically uniform emission and adding both directional features. The extrapolated result is \SI{778 \pm 147}{\micro\gram} and compares well with the modelled total value of \SI{613 \pm 83}{\micro\gram} within the margins of uncertainty.

The observation of a larger quantity of directional debris towards the target normal on the side of the laser interaction is similar to an earlier observation \cite{Booth2018}, pointing to small and fast projectiles which are emitted in the early phase of the interaction. The directional emission of dense flares of debris is favourable in situations of tight laser focusing. The latter is required to reach ultra-high intensities, but brings the precious final focusing optic into close vicinity to the debris source. As larger fragments of slow debris are likely emitted spherically uniform for late times, they pose a much lower risk as they can be addressed by available mitigation schemes, i.e. spinning protection disks \cite{Chen2024}. The observed small asymmetry of the spherical emission (with more deposition towards the target rear side) might be owed to the asymmetry of the charge distribution in the environment of the laser-interaction, with more electrons deposited in laser forward direction.

The experimental results show two small directional marks next to one large mark, which is counter-intuitive when comparing with the modelling that predicts a larger total emission of debris for smaller targets \cite{Ehret2023}. If the presumption is correct that both small marks correspond to both shots on small targets, then the directional fraction of debris is smaller for small disk targets than for large ones. In conjunction with a likely early emission of this directional emission \cite{Booth2018}, this can be related to the increased flatness of the target potential when comparing large planar targets to smaller ones. Following this hypothesis, the limit case of a spherical potential would lead to a minimum amount of directional debris whereas a maximum amount would be issued by the potential of an infinite plate.

The characteristic hourglass shape of the directional debris marks might encode valuable information about the laser-target interaction. Studies on laser induced forward and backward transfer in the long-pulse regime show the ejection of debris dependent on laser pulse width, laser pulse energy density and target–catcher distance \cite{Sano2002, Papa1999, Hennig2012}. Further investigation is required to evaluate if debris can be an auxiliary metrology on laser focal spot profile and temporal laser contrast.

This work took benefit of uniform absorption curve of nickel across the visible spectrum to introduce a fast spatially resolved way of debris characterization. When using spectrometers instead of a flatbed scanner, surface plasmons might be a way to characterize not only the thickness of a layer but also the size of nano-structures, when using materials that exhibit a large surface plasmon strength \cite{Rycenga2011}.

\section{Conclusion}\label{sec:conclusion}

We present a novel method for the characterization of thin layers of debris deposit based on RGB transmission scans that can be performed with commercial flatbed scanners. Initially transparent debris shields from fused silica are successfully used as debris catchers during experiments with high-power ultra-relativistic laser-pulses irradiating solid density targets. Scans reveal two distinct types of debris: (i) \SI{\approx 3}{\percent} is directed in narrow emission cones away from target front- and rear-side normal direction, and (ii) \SI{\approx 97}{\percent} are emitted spherically. While more debris of type (i) is emitted away from the target front, type (ii) shows a slight asymmetry favouring the target rear side. The former agrees with previous works \cite{Booth2018}, the latter might be due to the overall asymmetric space charge distribution induced by the laser-plasma interaction.

The quantitative characterization of the amount of debris and the direction of ejection can be used to promote the implementation of novel schemes that mitigate its deleterious effect on optical components and diagnostics.

\section*{Author Contributions}

The author contributions are as follows: ME commissioned the methodology; CV, AH acquired the data; IMV performed the data curation and analysis; IMV wrote the first draft of the manuscript; JLD, IMV, ME revised the methodology; PWB, ME organized the beamtime at ELI NP; SA, HA, JIA, MC, JLD, MG, MK, RL, DL, AM, DS, DU were involved with underlying experimental work; all authors contributed to manuscript improvement, read, and approved the submitted version.

\begin{acknowledgments}

This work would not have been possible without the help of the laser- and the engineering teams at CLPU and ELI NP. Special thanks for much appreciated support to the workshops of CLPU and ELI NP. The authors are grateful to D.~de~Luis and to P.~Puyuelo Vald\'es for help in preparing the experimental campaign. This work received funding from the European Union’s Horizon 2020 research and innovation program through the European IMPULSE project under grant agreement No. 871161; from grant PDC2021-120933-I00 funded by MCIN/ AEI / 10.13039/501100011033 and by the European Union NextGenerationEU/PRTR; from grant PID2021-125389OA-I00 funded by MCIN / AEI / 10.13039/501100011033 / FEDER, UE and by "ERDF A way of making Europe", by the European Union and in addition from Unidad de Investigación Consolidada de la Junta de Castilla y León  No. CLP087U16. UPM47 campaign was funded through IOSIN, Nucleu PN- IFIN-HH 23-26 Code PN 23 21, and the Extreme Light Infrastructure - Nuclear Physics (ELI-NP) Phase II, a project co-financed by the Romanian Government and the European Union through the European Regional Development Fund and the Competitiveness Operational Programme (1/07.07.2016, COP, ID 1334). This research was funded, in part, by the French Agence Nationale de la Recherche (ANR), Project No. ANR-22-CE30-0044. 

\end{acknowledgments}

\section*{Data Availability Statement}

The raw data and numerical methods that support the findings of this study are available from the corresponding author upon reasonable request.

\appendix{}
\renewcommand{\thesection}{\large \Alph{section}}

\begin{table}[!b]
\caption{\label{tab:calibrationBC}%
Grayscale to OD calibration fit parameters for every color channel of an EPSON V-750-PRO.}
\begin{ruledtabular}
    \begin{tabular}{ccc}
          & B & C  \\
         \midrule
          Red & \num{1.333 \pm 0.011} & \num{8.311 \pm 0.067} \\  
          Green & \num{1.351 \pm 0.011} & \num{8.199 \pm 0.067} \\
          Blue & \num{1.471 \pm 0.011} & \num{7.532 \pm 0.067} \\ 
    \end{tabular}
\end{ruledtabular}
\end{table}

\section{\large Scanning procedure}
\label{app:scans}
\begin{figure}[!tbp]
    \centering
    \includegraphics[width=\columnwidth,trim=0.4cm 0.4cm 0.4cm 0.4cm,clip]{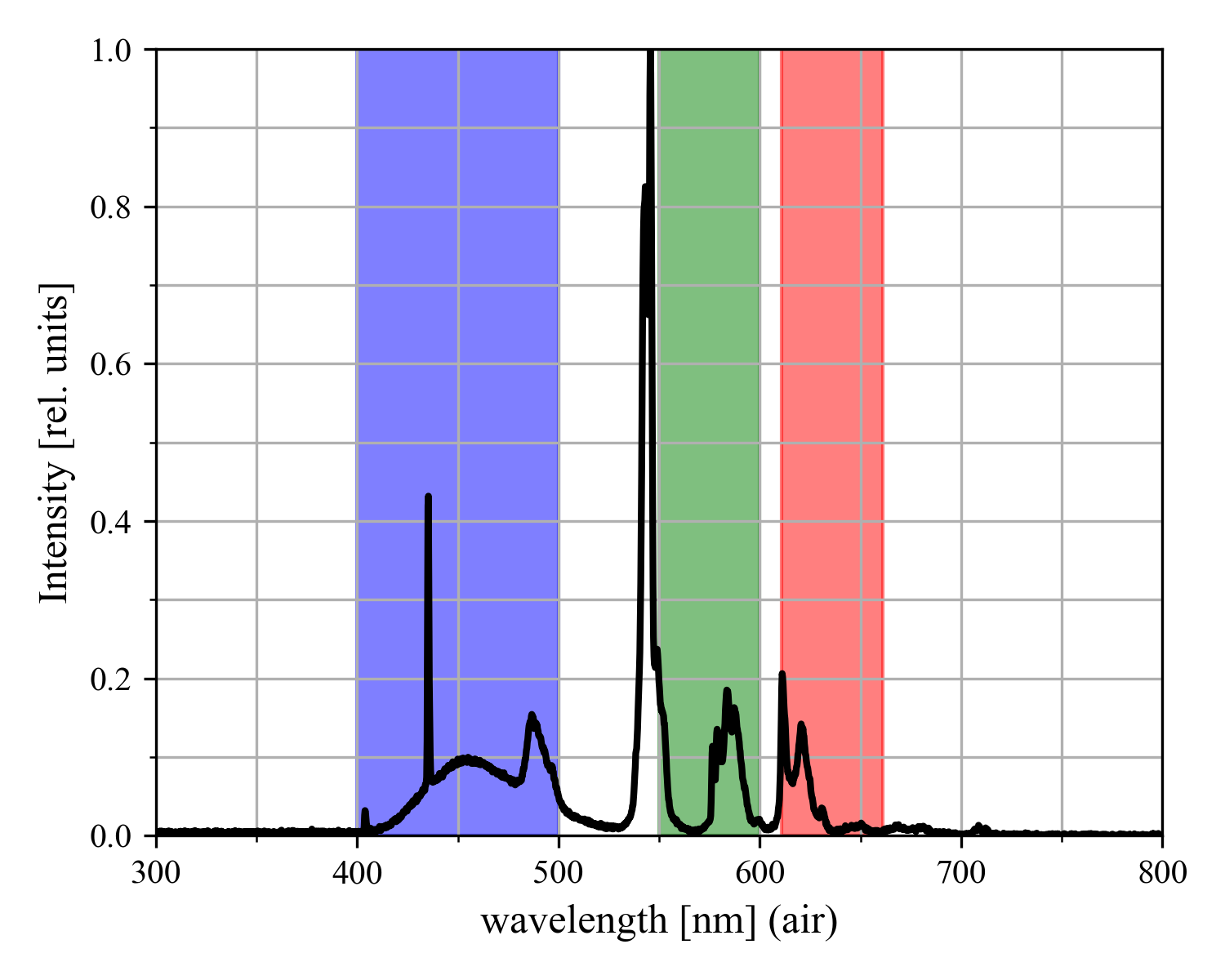}
    \caption{Normalized spectrum of the light source of the EPSON V-750-PRO flatbed scanner used for this work; with an indication of the blue, green and red bands of acquisition for the scanner head.}
    \label{fig:EPSON}
\end{figure}

Scans with an EPSON V-750-PRO flatbed scanner are performed in both possible orientations (with the debris facing the scanner light source and with the debris facing the scanner readout) and differences are taken into account as uncertainty of the intensity $\Delta I$. To improve the estimate of uncertainties of the presented method, RGB-color scans are performed and results of the differing acquisition bands are compared (B~:~\SIrange{400}{500}{\nano\metre}; G~:~\SIrange{550}{600}{\nano\metre}; R~:~\SIrange{611}{661}{\nano\metre} \cite{Larraga2018}). The spectrum $s(\lambda)$ of the scanner lamp is compared to the acquisition bands in Fig.~\ref{fig:EPSON}. Here $\lambda = 2\pi c / \omega$ is the vacuum wavelength, with the speed of light $c$ and the frequency of the electromagnetic wave $\omega$.

The transmittance $T = \int_{\omega_-}^{\omega_+} \mathfrak{T}(\omega) \cdot s(\omega) ~ \mathrm{d}\omega$ can be derived from the optical density (OD) of a sample. The OD detected in a pixel of coordinates $(m,n)$ is defined as the logarithmic ratio between the transmitted grayscale intensity $I(m, n)$ and the scanner response for a scan without sample $I_0$, which can be written as
\begin{equation}
    \mathrm{OD} = - \log_{10}{ \left[ \frac{I(m,n)}{I_0} \right] } = - \log_{10}{ \left[ T_{mn} \right] } \quad .
    \label{eq:OdTrasnmittance}
\end{equation}
Kodak WRATTEN 2 Neutral Density No. 96 Filters with well defined spectral properties are used for absolute scanner calibration of transmission scans. The relation of optical density to the transmitted grayscale intensity results to
\begin{equation}
    \mathrm{OD} = C - \frac{\ln{ \left[ I_{mn} \right]}}{B}  \quad ,
    \label{eq:OdSignal}
\end{equation}
where $B$ and $C$ are band-dependent fit parameters shown in Tab.~\ref{tab:calibrationBC}. $B$ represents the inverse scale parameter for the exponential decay and $C$ is the minimum detectable transmittance. The dynamic range on the grayscale can be computed as $\exp{\left[ B \cdot C \right]}$. One retrieves the transmittance as
\begin{equation}
     T_{mn} = 10^{ \left( \ln{\left[ I_{mn} \right]} / B \right) - C } \quad ,
    \label{eq:Trasnmittance}
\end{equation}
with a relative uncertainty of
\begin{align}
     &\frac{\Delta T_{mn}}{T_{mn}} = \ln{\left[ 10 \right]} \nonumber \\
     &\cdot \sqrt{ \left( \frac{\Delta I_{mn}}{I_{mn} B} \right)^2 + \left( \frac{ \ln{\left[ I_{mn} \right]} \Delta B}{B^2} \right)^2 + \left( \Delta C \right)^2}  \quad .
    \label{eq:DeltaTrasnmittance}
\end{align}

The relative uncertainty calculates to $\Delta T_{mn} / T_{mn} \approx \SI{ 16}{\percent}$ for all color channels on the scale of 16-bit images used for this work, considering a scan-to-scan uncertainty of $\Delta I_{mn} / I_{mn} \approx \SI{1.1}{\percent}$.

\section{\large Transmission of a flat double-layer system of debris and catcher}
\label{app:transmission}

For \si{\nano\metre}-scale layers of metal deposits on transparent support plates it is possible to retrieve the debris thickness from a measurement of the transmittance
\begin{equation}
    T(z_\textrm{M},z_\textrm{S}) = \int_{\omega_-}^{\omega_+} \lVert \mathfrak{T}(z_\textrm{M},z_\textrm{S},\omega) \rVert^2 \cdot s(\omega) ~ \mathrm{d}\omega \quad ,
    \label{eq:transmittance}
\end{equation}
where $z_\mathrm{M}$ is the thickness of the metal deposit, $z_\mathrm{S}$ is the thickness of the support plate, and $\omega$ denotes the frequency of the incident electromagnetic waves in a normalized spectrum $\int_{\omega_-}^{\omega_+} s(\omega) ~ \mathrm{d}\omega = 1$. The transmission ratio $\mathfrak{T}$ for a mono-chromatic incident wave is defined as
\begin{equation}
    \mathfrak{T}(z_\textrm{M},z_\textrm{S},\omega) = E_\mathrm{t} / E_0^+
    \label{eq:transmissionratio}
\end{equation}
with the electric field amplitude of the incident wave $E_0^+$ and the amplitude at the exit of the double layer system $E_\mathrm{t}$. For normally incident electromagnetic fields, the continuity of electric- and magnetic-field across interfaces between layers of media $j$ and $j+1$ implies at the boundary
\begin{align}
    \mathfrak{E}_j^+ + \mathfrak{E}_j^- &= \mathfrak{E}_{j+1}^+ + \mathfrak{E}_{j+1}^- \nonumber \\
    \frac{1}{\eta_j} \left( \mathfrak{E}_j^+ - \mathfrak{E}_j^- \right) &= \frac{1}{\eta_{j+1}} \left( \mathfrak{E}_{j+1}^+ - \mathfrak{E}_{j+1}^- \right) \quad ,
    \label{eq:continuity}
\end{align}
with electric resistance $\eta_j = \sqrt{\mu_j/\epsilon_j}$, where $\mu_j$ is the permeability and $\epsilon_j$ the permittivity of the respective material. Here the electric field component is described in terms of plane waves $\mathfrak{E}_j^\pm = E_j^\pm \mathrm{e}^{\mp i ( z' k_j - \omega t ) }$ with wavenumber $k_j$. Components $\mathfrak{E}_j^+$ are forward propagating (in direction from $j$ to $j+1$) while components $\mathfrak{E}_j^-$ propagate backwards. One obtains

\begin{align}
    E_0^+ + E_0^- &= E_\mathrm{M}^+ + E_\mathrm{M}^- \nonumber \\[10pt]
    \frac{1}{\eta_0} \left( E_0^+ - E_0^- \right) &= \frac{1}{\eta_\mathrm{M}} \left( E_\mathrm{M}^+ - E_\mathrm{M}^- \right) \quad ,  \label{eq:interface1} \\[10pt]
   \sum_{p \in \{ +1, -1 \}} E_\mathrm{M}^p \mathrm{e}^{- i p z_\mathrm{M} k_\mathrm{M} }  &= E_\mathrm{S}^+ + E_\mathrm{S}^- \nonumber \\
     \sum_{p \in \{ +1, -1 \}} \frac{p}{\eta_\mathrm{M}} E_\mathrm{M}^p \mathrm{e}^{- i p z_\mathrm{M} k_\mathrm{M} } &= \frac{1}{\eta_\mathrm{S}} \left( E_\mathrm{S}^+ - E_\mathrm{S}^- \right) \quad ,  \label{eq:interface2}
\end{align}

\begin{align}
    \sum_{p \in \{ +1, -1 \}} E_\mathrm{S}^p \mathrm{e}^{- i p z_\mathrm{S} k_\mathrm{S} }  &= E_\mathrm{t} \nonumber \\
     \sum_{p \in \{ +1, -1 \}} \frac{p}{\eta_\mathrm{S}} E_\mathrm{S}^p \mathrm{e}^{- i p z_\mathrm{S} k_\mathrm{M} } &= \frac{1}{\eta_\mathrm{0}}  E_\mathrm{t} \quad ,  \label{eq:interface3}
\end{align}
at the entrance (Eqs.~\ref{eq:interface1}), middle interface (Eqs.~\ref{eq:interface2}) and exit (Eqs.~\ref{eq:interface3}) of the double layer system, where $E^-_0$ denotes the reflected wave.

For a first layer of the ferromagnetic transition metal nickel (Ni) followed by a second layer of Fused Silica (FS) one derives
\begin{widetext}
\begin{align}\label{eq:transmissionratio_solved}
     \mathfrak{T}(z_\textrm{Ni},z_\textrm{FS},\omega) =& \frac{8 \mathrm{e}^{- i z_\textrm{FS} k_\textrm{FS}}}{m_+ \mathrm{e}^{ i z_\textrm{Ni} k_\textrm{Ni}} \left( n_+ u_+  + n_- u_- \mathrm{e}^{ -2i z_\textrm{FS} k_\textrm{FS} } \right) + m_- \mathrm{e}^{-i z_\textrm{Ni} k_\textrm{Ni}} \left( n_- u_+ + n_+ u_- \mathrm{e}^{- 2i  z_\textrm{FS} k_\textrm{FS} } \right) }\\
    \Vert ~ \text{with} \quad m_\pm &= 1 \pm n_\textrm{Ni} \\
                              n_\pm &= 1 \pm \frac{n_\textrm{FS}}{n_\textrm{Ni}} \\
                              u_\pm &= 1 \pm \frac{1}{n_\textrm{FS}} \\
                              k_j   &= \frac{ n_j \omega}{c} - i \frac{\alpha_j}{2}
                              \label{eq:transmissionratio_k}
\end{align}
\end{widetext}
\noindent where $\alpha_j$ is the absorption coefficient of layer $j$, and $n_j$ denotes the refractive index respectively. The following approximations of both spectrally resolved quantities are evaluated for wavelengths in the range from \SIrange{400}{661}{\nano\metre} (from $\omega =$~\SIrange{4.71e15}{2.85e15}{\per\second} respectively).

The absorption of films of evaporated nickel is \cite{Johnson1974}
\begin{equation}
    \alpha_\mathrm{Ni} \approx \left( \frac{\omega}{\SI{1.64 \pm 0.05 e18}{\per\second}} + 0.08017 \right) \si{\per\nano\metre} \quad ,
\end{equation}
such a monotonic behaviour is common for transition metals that do not build surface plasmons efficiently \cite{Axelevitch2012}. Note that the absorption coefficient $\alpha_\mathrm{FS}$ of silica glass is neglected in the following for its small magnitude \cite{Kitamura2007}. The refractive index of nickel \cite{Johnson1974} and Fused Silica \cite{Malitson1965} are
\begin{align}
    n_\mathrm{Ni} &\approx \frac{\SI{2.43 \pm 0.08 e15}{\per\second}}{\omega} + 1.183 \quad , \\
    n_\mathrm{FS} &\approx \left( \frac{\omega}{\SI{3.18 \pm 0.05 e16}{\per\second}} \right) ^2 + 1.448 \quad .
\end{align}

The evaluation of Eq.~\ref{eq:transmittance} can now be performed numerically in spectral slices, i.e. for every color channel of a scan (compare Fig.~\ref{fig:EPSON} for the spectrum and the acquisition bands of an EPSON V-750-PRO flatbed scanner). Note that some application cases might be well fitted with an approximated analytical solution for a thin film on a thick finite transparent substrate \cite{Swanepoel1983}.

\end{document}